%% file: main.tex
\documentclass[journal]{vgtc}                
\ifpdf
  \pdfoutput=1\relax                   
  \usepackage{graphicx}                
  \DeclareGraphicsExtensions{.pdf,.png,.jpg,.jpeg} 
\else
  \usepackage{graphicx}                
  \DeclareGraphicsExtensions{.eps}     
\fi%

\graphicspath{{figures/}{pictures/}{images/}{./}} 

\usepackage{microtype}                 
\PassOptionsToPackage{warn}{textcomp}  
\usepackage{textcomp}                  
\usepackage{mathptmx}                  
\usepackage{times}                     
\usepackage{cite}                      
\usepackage{tabu}                      
\usepackage{booktabs}                  

\usepackage{amsmath}
\usepackage{algorithm}
\usepackage[noend]{algpseudocode}

\usepackage{color}
\usepackage{paralist}
\usepackage{enumitem}
            
\newcommand{\etal}{{\it et al.\ }}

\definecolor{bluecrayola}{rgb}{0.12,0.46,1.0}

\newcommand{\copied}[1]{{\color{gray}\textit{[Placeholder: #1]}}}

\newcommand{\tool}{Data Player~}
\newcommand{\toole}{Data Player}

\definecolor{mygreen}{RGB}{0,176,80}
\definecolor{mygreen2}{RGB}{232, 245, 233}
\definecolor{myorange}{RGB}{192, 0, 0}
\definecolor{myblue}{RGB}{47, 85, 151}
\newcommand{\highlightone}[1]{\colorbox{mygreen2}{\textcolor{mygreen}{#1}}}
\newcommand{\highlighttwo}[1]{\colorbox{myblue}{\textcolor{white}{#1}}}

\newcommand{\revise}[1]{{\color{black} {#1}}}
\newcommand{\review}[1]{\textcolor{purple}{}}

\onlineid{1049}

\vgtccategory{Research}

\title{\toole: Automatic Generation of Data Videos\\ with Narration-Animation Interplay}

\author{Leixian Shen, Yizhi Zhang, Haidong Zhang, and Yun Wang}
\authorfooter{

\item
L. Shen is with The Hong Kong University of Science and Technology.
E-mail: lshenaj@connect.ust.hk.
\item
Y. Zhang is with Cornell University. E-mail: yz2668@cornell.edu.
\item
H. Zhang and Y. Wang are with Microsoft Research Asia (MSRA).
\\E-mail: $\{haizhang, wangyun\}$@microsoft.com. 
\\Y. Wang is the corresponding author.
\item
Work done during L. Shen and Y. Zhang’s internship at MSRA.

}

\shortauthortitle{Biv \MakeLowercase{\textit{et al.}}: Global Illumination for Fun and Profit}

\abstract{
\input{sections/abstract}
} 


\CCScatlist{ 
 \CCScat{K.6.1}{Management of Computing and Information Systems}%
{Project and People Management}{Life Cycle};
 \CCScat{K.7.m}{The Computing Profession}{Miscellaneous}{Ethics}
}

\teaser{
\centering
\includegraphics[width=\linewidth]{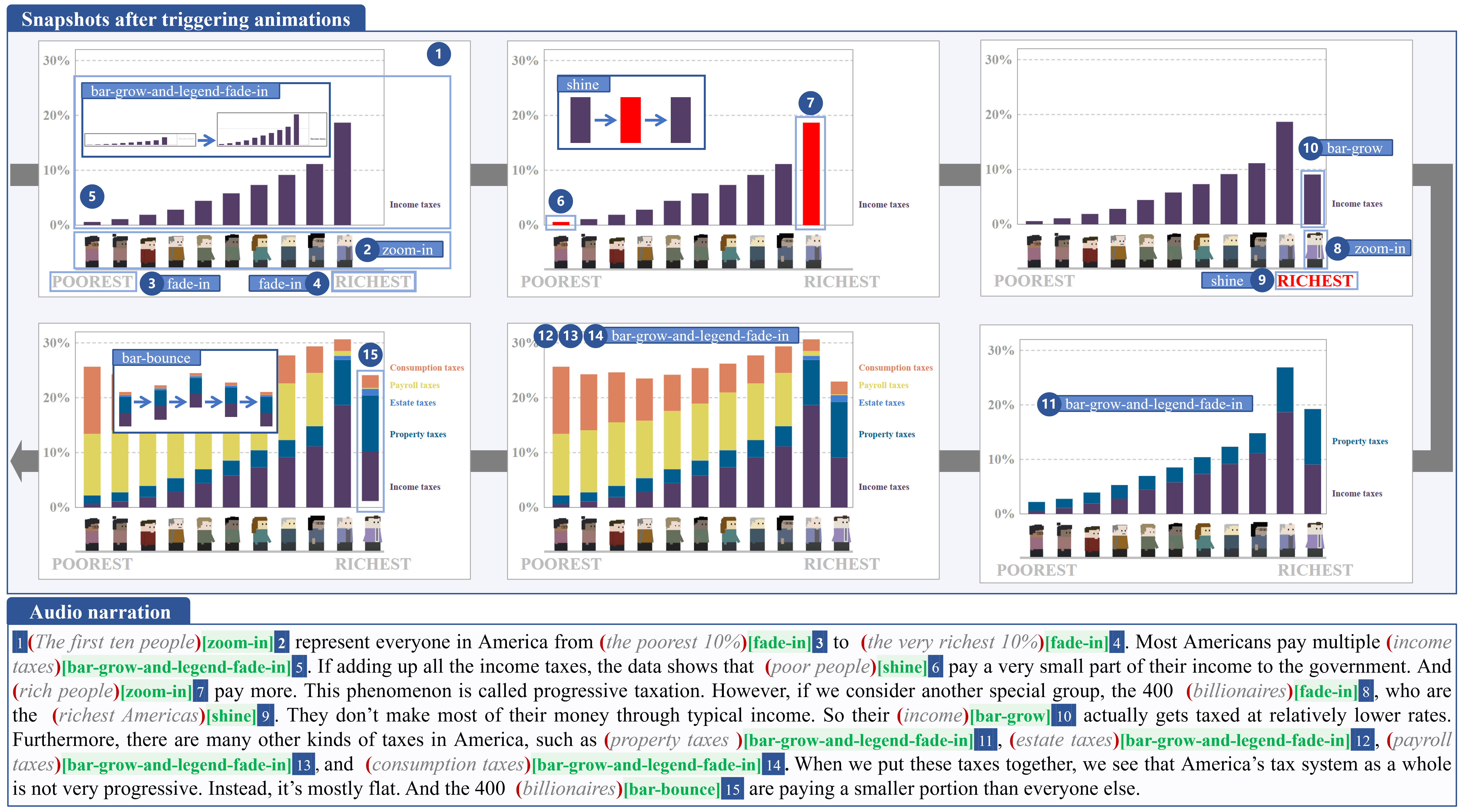}
\caption{A data video example with narration-animation interplay automatically generated by \tool from a static visualization and descriptive text, where 
\textcolor{myorange}{\textbf{(}}\textcolor{Gray}{\textit{narration segment}}\textcolor{myorange}{\textbf{)}}\highlightone{{\small animation-effect}}\highlighttwo{{\small x}} means the animation effect will be triggered when the audio reaches the corresponding narration segment and lasts for the duration of the segment span in the audio, resulting in the $x$-th keyframe above.
 }
\label{fig: teaser}
}

\begin{document}
\maketitle

\input{sections/introduction}
\input{sections/related-work}

\input{sections/formative-study}
\input{sections/approach}
\input{sections/user-study}

\input{sections/discussion}
\input{sections/conclusion}
\bibliographystyle{abbrv-doi}

\bibliography{references}

\clearpage
\end{document}

%% file: sections/abstract.tex
Data visualizations and narratives are often integrated to convey data stories effectively. Among various data storytelling formats, data videos have been garnering increasing attention. These videos provide an intuitive interpretation of data charts while vividly articulating the underlying data insights. However, the production of data videos demands a diverse set of professional skills and considerable manual labor, including understanding narratives, linking visual elements with narration segments, designing and crafting animations, recording audio narrations, and synchronizing audio with visual animations.
To simplify this process, our paper introduces a novel method, referred to as \toole, capable of automatically generating dynamic data videos with narration-animation interplay. This approach lowers the technical barriers associated with creating data videos rich in narration. 
To enable narration-animation interplay, \tool constructs references between visualizations and text input. 
Specifically, it first extracts data into tables from the visualizations. Subsequently, it utilizes large language models to form semantic connections between text and visuals. Finally, \tool encodes animation design knowledge as computational low-level constraints, allowing for the recommendation of suitable animation presets that align with the audio narration produced by text-to-speech technologies.
We assessed \toole's efficacy through an example gallery, a user study, and expert interviews. The evaluation results demonstrated that \tool can generate high-quality data videos that are comparable to human-composed ones. 

%% file: sections/introduction.tex
\section{Introduction}


Visualization and corresponding descriptions often work together for data storytelling. Combining data visualization and narratives, data videos have become popular among practitioners as a visual storytelling form in fields such as journalism, marketing, and education~\cite{segel2010narrative,amini_understanding_2015}.
Over more than a decade of research, data videos have demonstrated their ability to deliver condensed information, increase audience engagement, and support comprehension and memorization of data facts in data communication~\cite{wang2022investigating,amini_hooked_2018,shi2021communicating}.

In a data video, \review{R2.1.1}\revise{rich information is usually packed compactly and delivered through the coordination of audio narrations and animated graphics. As indicated by the dual-coding theory, human cognition can process verbal and visual objects simultaneously, and both of them play an essential role~\cite{clark_dual_1991}.} 
However, creating such a data video requires a variety of skills in multiple areas, including \review{S.1}\revise{understanding narratives}, visual animation design, narration scripting, and time alignment of audio and animations, which are usually difficult to perform for novices without instruction. 

To help users overcome these barriers, various technologies have been developed for different aspects of data video creation.
For example, the visualization community has developed data visualization animation-specific technologies to facilitate the animation creation process, such as declarative specification grammars~\cite{ge_canis_2020, kim_gemini_2021, Zong2022}, authoring systems~\cite{ge_cast_2021,thompson2021data,amini2017authoring}, and automated algorithms~\cite{wang_animated_2021,kim_gemini2_2021,shi_autoclips_2021}. However, they neglect the importance of narration-animation interplay.
In a recent study, Cheng et al.~\cite{wang2022investigating} investigated the role of narrations and animations in data videos. They found that users usually have static visual designs and text descriptions at hand for storytelling, and narration-animation interplay can effectively enhance liveliness compared to static forms.
There are also a set of works that link static text and visualization together, using text-visual interplay to enhance readability in the form of interactive documents~\cite{latif2021kori, Charagraph}, visualization annotations~\cite{lai_automatic_2020}, etc.
However, they do not fully exploit the potential of data animation to model how the data changes over time or space.
In conclusion, existing authoring tools lack features for integrating narration with data animations in data videos for engaging storytelling.

To address this gap, this paper targets to design an intuitive and powerful approach that enables the automatic creation of informative data videos with narration-animation interplay from static visualizations and accompanying descriptive text. 
To achieve this, we first conducted a formative study to understand users' process of crafting data videos and explore the key design considerations of data video creation in their prior experience. From the study, we derived a set of high-level design constraints. The interviewees also expressed the need for support in \review{S.1}\revise{understanding narratives}, linking text and visuals, generating animations, and aligning the timeline of audio and animations. 



In response to the feedback, we take the first step towards automating the generation of data videos with narration-animation interplay and design \toole, which automates the four stages above, lowering the technical barriers of creating data videos, especially for novices. 
To enable narration-animation interplay, \tool constructs references between visualizations and text input. We first extract data from input visualizations so that we convert the text-visual linking challenge into a matching problem between data table rows and narration segments. Subsequently, \tool leverages the powerful natural language understanding ability of Large Language Models (LLMs) to associate narrative words with related data table rows, thereby establishing links between textual and visual elements. 
\tool further produces a sequence of animation by modeling animation design as a Constraint-Satisfaction Problem (CSP). In detail, text-to-speech technologies are adopted to automatically generate audio narrations with timestamps of each word. \tool then encodes design knowledge learned from the formative study and existing literature into computational low-level constraints, which are further fed into the constraint solver to generate suitable animation sequence with a pre-defined animation library. Finally, the audio and animations are rendered into a data video with narration-animation interplay. 

To evaluate the liveliness of data videos produced by \toole, we curated an example gallery and conducted a user study and an expert interview. The results showed that the automatic-generated data videos are comparable to the human-composed ones, suggesting that \tool can effectively produce data videos with narration-animation interplay, conveying the intended information while engaging the audience.
The main contributions of this paper are as follows:
\begin{compactitem}
\item
A formative study to understand users’ processes and key design considerations of data video creation, leading to a set of high-level design constraints for the automatic coordination of narration and animation in data videos.

\item
\toole, an innovative approach that takes the first step towards the automatic generation of vivid data videos with narration-animation interplay from a static visualization and its description. \tool leverages LLMs to \review{S.1}\revise{understand narratives} and establish text-visual links. It further uses constraint programming to recommend suitable animation sequences.




\item
An example gallery, a user study, and an expert interview to evaluate the effectiveness of \toole. The results demonstrated that \tool can automatically produce human-comparable data videos with narration-animation interplay.

\end{compactitem}

%% file: sections/related-work.tex
\section{Related Work}
\tool draws upon prior efforts in data video creation, visualization-text interplay, and constraint-based generation approach.
\subsection{Data Videos Creation}
Data video is one of creative data presentation genres~\cite{segel2010narrative}, which uses animations and audio in addition to static data to provide additional channels of communication for information transformation~\cite{amini_hooked_2018}. 

Prior studies have contributed insights into the comprehension and creation of animated data visualizations. 
For example, Tversky et al.~\cite{tversky_animation_2002} first suggested two design principles, i.e., Congruence and Apprehension, which was followed by Heer and Robertson~\cite{heer_animated_2007}, who proposed ten specific guidelines that focus on animated transitions in statistical data visualizations based on the two initial overarching principles.
Amini et al.~\cite{amini_understanding_2015} conducted a systematic analysis of 50 data videos, identifying the most commonly used visualization types and attention cues, as well as high-level narrative structures. Their findings confirmed that animation in data videos has a positive effect on viewer engagement~\cite{amini_hooked_2018}.
Thompson et al.~\cite{thompson_understanding_2020} analyzed design primitives of animated data-driven graphics from four perspectives: object, graphics, data, and timing.
Further examining the animated visual narratives in data videos, 
Shi et al.~\cite{shi2021communicating} developed a design space for motion design. 

Furthermore, numerous authoring and programming tools have been created and are being continually developed to facilitate the production of animations and data videos. These tools are intended to enable creators to bring their ideas to life in a more efficient and effective manner~\cite{Chen2022}. 
General video creation tools (e.g., Adobe After Effects and Premier) provide fine-grained control of videos, but they require a high level of expertise and manual effort and are not tailor-made for data videos.
In the visualization community, animation-specific grammars (e.g., Canis~\cite{ge_canis_2020}, Gemini~\cite{kim_gemini_2021}, and Animated Vega-Lite~\cite{Zong2022}) have been developed to provide high-level specification languages for implementing keyframe-based animated transitions in data graphics. However, it requires programming skills and can be laborious for the authors.
To ease the process, interactive user interfaces have emerged to enable novices to create their own data videos. DataClips~\cite{amini2017authoring} allows novices to create data videos by selecting and concatenating pre-defined animations from a comprehensive library, which includes clips that are suitable for visualizing different types of data. Based on the library, Kineticharts~\cite{lan2021kineticharts} enhances users' emotional engagement by improving the storytelling aspect of data presentation without compromising users' comprehension of the data.
CAST~\cite{ge_cast_2021} and Data Animator~\cite{thompson2021data} support recommendations for auto-completion so that users only need to provide keyframes. 
Researchers have also developed automatic approaches to further reduce time-consuming manual operations.
Gemini2~\cite{kim_gemini2_2021} improves Gemini~\cite{kim_gemini_2021} by providing keyframe suggestions to help users create visually appealing animations. 
InfoMotion\cite{wang_animated_2021} enables the recommendation of animations of infographics based on their graphical properties and information structures.
AutoClips~\cite{shi_autoclips_2021} allows users to easily input a sequence of data facts, which are then automatically transformed into a polished data video. 

However, these works have neglected an important channel, narration, when applying these techniques to data videos. Cheng \etal~\cite{wang2022investigating} recently investigated the role and interplay of narrations and animations and identified close links between the two perspectives. Following the study, our work serves as the first step towards the automatic generation of data videos with narration-animation interplay.

\subsection{Visualization-Text Interplay}
The interplay between visualization and text plays an important role in data storytelling~\cite{shen2022towards}. 
Recent studies have shown that the separation of text and charts may cause a split-attention effect and introduce cognitive burden for users~\cite{latif2021kori}. By contrast, linking visualization and text can promote the communication of data facts~\cite{srinivasan_augmenting_2019}, support the interpretation of machine learning models~\cite{hohman_telegam_2019}, and enhance readers' comprehension and engagement~\cite{latif2021kori,zhi_linking_2019}.

Given these benefits, researchers have actively integrated visualizations and text for interactive purposes in data presentations. For example, Vis-Annotator~\cite{lai_automatic_2020} automatically presents annotated charts according to the text description. 
Kong et al.~\cite{kong2014extracting} proposed a crowdsourcing method to collect high-quality annotations for the references of charts and text. Subsequently, automatic techniques are proposed to link text and charts with rule-based algorithms~\cite{metoyer_coupling_2018}. 
Latif et al.~\cite{latif2021kori} further proposed a framework to construct references between text and charts in data documents by explicitly declaring the links. 
The study from Kim et al.~\cite{kim_facilitating_2018} found that text-table linking in documents can support readers to pursue content better with highlighted cells. And the interactive data articles enhanced with widgets such as ``stepper'' and ``scroller'' also enable the control for users to be more autonomous during their reading~\cite{zhi_linking_2019,mckenna_visual_2017}.
\review{R3.5}\revise{In addition, CrossData~\cite{Chen2022a} leverages text-data links to interactively author data documents. 
}
To further ease the process of creating text-chart connections and support chart highlighting, the following studies have developed programming language~\cite{conlen_idyll_2018}, authoring tools~\cite{sultanum_leveraging_2021}, and interactive approaches~\cite{Charagraph, DataParticles2023}.

Different from static charts and documents, the dynamic changes with time progressing in videos grant it its own narrative structures~\cite{segel2010narrative}. Hence, the visualization-text linking in static data stories needs to be extended to narration-animation interplay in data videos~\cite{wang2022investigating}. 
To further unleash the power of integrating oral narration and visual animation in data videos, our work steps towards the automatic transformation of static text and visualizations into engaging data videos with narration-animation interplay.

\subsection{Constraint-Based Generation Approach}
Constraint-based approaches have been widely applied to generate visualizations~\cite{moritz_formalizing_2019,shendata}, interface alternatives~\cite{swearngin_scout_2020, xu_global_2014}, and short videos~\cite{chi_automatic_2020, Chi2021a}. 
For example, URL2Video~\cite{chi_automatic_2020} captures quality materials and design styles extracted from a web page and converts them into a short video given temporal and visual constraints. While not taking narratives into consideration and are not data-oriented, it inspires us to extract design elements from static visualizations and arrange them in the animation timeline based on pre-defined constraints. However, the scenario of data video introduces new challenges for the design of narration-animation interplay~\cite{wang2022investigating}. 
On the other hand, Moritz et al.~\cite{moritz_formalizing_2019} demonstrated that  theoretical design knowledge can be expressed in a concrete, extensible, and testable form by modeling them as a collection of constraints. Therefore, we adopt a constraint-based method to model our derived design knowledge about narration-animation interplay and incorporate them into an automatic creation workflow. The resulting approach can recommend data video designs satisfying different aspects of guidelines and further facilitate designers' crafting.

%% file: sections/formative-study.tex
\section{Formative Study} \label{sec: formative study}
The goals of the formative study are to (1) understand practitioners' process of video crafting, and (2) explore the key design considerations of narration-animation interplay in their previous design experiences.

\subsection{Participants}
To achieve the above goals, we recruited 10 participants from both academia and industry with diverse backgrounds, including professional designers of video, motion graphics, animation, film post-producer, and visualization researchers.
They have all acquired professional training or degrees, including three Ph.D.s, five M.S.s, and two B.S.s.
All of them have experience in data video crafting through professional tools (e,g., Adobe After Effects and Premiere) or other simplified video creation tools (e.g., Microsoft PowerPoint and iMovie), with a self-reporting level of familiarity with this area (\emph{M} = 4.12, \emph{SD} = 0.83, range = 3-5 with 1 = ‘‘No Experience’’ and 5 = ‘‘Expert’’). 
They were aged 24–32 years (5 females and 5 males, \emph{M} = 27, \emph{SD} = 3.10).
We recruited them through online advertising and word-of-mouth.  

\subsection{Study Setup}
The study procedure consists of two sessions with retrospective analysis and semi-structured interviews, respectively. 
First, 
we conducted a retrospective analysis, which has been proven to be an effective method for reconstructing participants' behaviors, rationales, and emotions for previous events \cite{russell_looking_2014}. Participants were asked to provide and show 2-3 examples from their prior data storytelling works. To promote reflection, they were required to demonstrate the creation process and explain the rationale for their design decisions.
Finally, 25 videos were presented, covering 8 common chart types (e.g., maps, bars, lines, etc.). 

After the retrospective analysis, we held one-on-one semi-structured interviews with the participants. The questions focused on concrete examples of narration-animation interplay in the works shown, allowing participants to recall more details of their designs and provide more useful information. We also explored the participants' views on design principles of narration-animation interplay by asking about the role of narratives and motions in their projects and the relationship between them. Finally, participants shared their difficulties in crafting data videos, particularly in aligning narrations and animations, providing insights for automatic workflows. The entire process lasted about 90 minutes and was recorded for subsequent analysis. The participants were compensated \$15 for their time.

\subsection{Data Video Creation Process}
Participants normally prepare materials including visualization vector graphics and narrative scripts before crafting data videos in professional software such as Adobe After Effects, Cinema 4D, and Blender. 
During this preparation phase, they focus on the aesthetic design of their graphics and descriptive narrative writing, with little attention to the dynamic interplay between them and the motion effects of output videos. 
In some cases, graphics and texts may also be given to the creator by other collaborators such as graphic participants or screenwriters. After that, our findings identified four distinct design stages in terms of video crafting, which is of interest to our research. 

\textbf{Stage 1: Refining Narration Text.}
At this stage, participants try to collect the text descriptions for the charts and compose the narration text of the video to produce audio narration. 
If the text description is not created by them, they need to understand the intents that the text authors or storytellers would like to convey to the audience. In this stage, data video creators usually decompose the messages in the text narration and formulate the messages to convey to prepare for their further design of animation in the videos. 

\textbf{Stage 2: Building Visual References.}
Based on the formulated messages, participants match them to the visual references in the visualizations. Visual references are graphic elements in charts, such as the specific line in a line chart or the related rectangle in a bar chart. Participants build visual references associated with the messages in the narratives. Sometimes they may group them if there are some relationships between the visual elements. For example, one interviewee told us that he usually grouped two comparative data elements so that he designs animations for them in the subsequent motion design phase.

\textbf{Stage 3: Animation Design with Semantic Metaphors.}
After preparing the text narration and visualizations, participants design animations with semantic metaphors in line with the intent of storytelling. For example, if the storyteller wants to express a rising trend for a line chart, participants would make a dynamic growth curve from the lowest point to the highest point. 
When emphasizing the visual elements, participants may modify transparency, saturation, contours, or other visual properties as animation cues to highlight information. 
\review{R1.3}\revise{Additionally, three types of animation effects are most commonly adopted: entering, exiting, and emphasizing.}

\textbf{Stage 4: Coordination of Audio Narration and Animations.}
Finally, participants align the created series of animations with the audio created by the voiceover artist on the timeline. They often take the approach of manually adjusting the keyframes, setting the start frame and the end frame of the animation at the corresponding time points in the narrative audio. Most interviewees (8/10) reported that this process was very time-consuming and laborious because they needed to listen to the audio and watch the animation repeatedly to check for out-sync.

\subsection{Design Constraints}\label{sec: design constraints}
Through retrospective analysis and interviews, we found that participants were concerned about the echoes of narratives and animations in four dimensions: visual structure, data facts, semantics, and temporality.
Further, we derived a set of design constraints from interviewees' considerations collected from the study, as well as existing literature~\cite{wang2022investigating,amini_hooked_2018,shi2021communicating,tversky_animation_2002, heer_animated_2007}. 

First, the participants emphasized the importance of visual constraints in creating an organized and logical presentation. They suggested grouping relevant visual elements associated with the same data fact, establishing a sense of hierarchy and facilitating audience comprehension. Additionally, they recommended introducing unrelated background elements, such as titles and axes, at the beginning of the video to provide context. Maintaining consistent animation effects within groups of similar elements further enhances clarity.

The participants mentioned the importance of data interplay in enhancing audience understanding. They advocated for echoing narrations and animations by reiterating conveyed information, as well as selecting and animating visual elements relevant to the data facts in the narrative script. This approach helps emphasize key points and maintain the audience's attention. Moreover, incorporating animations into data facts, rather than other narratives, can improve comprehension. 

\review{R2.1.3}\revise{The participants also highlighted the role of semantic rules in conveying the intended message accurately, which refers to the implicit interactions that arise from the meanings and intentions of visual elements and their narrations in voiceover.}
They underscored the need to align the semantic intents of narrations and animations to avoid confusion. Furthermore, they suggested supplementing missing narrative information with annotations, such as labels, explanations, and context, to provide a more comprehensive representation of data.

Finally, temporal interplay emerged as another critical aspect. The participants stressed the importance of coordinating animation sequences with narrative structures to preserve the meaning of the information, synchronizing narrations and animations for consistency, and adapting the timing of animations to match narrations, ensuring that the information remains manageable for the audience.

We regard these design constraints as high-level because they cannot be directly translated into actionable computational programs, but lay the foundation for the subsequent formalization of low-level design constraints, which will be discussed in detail in Section~\ref{sec: Constraint Encoding}.

%% file: sections/approach.tex
\begin{figure*}[t]
\centering
\includegraphics[width=0.88\linewidth]{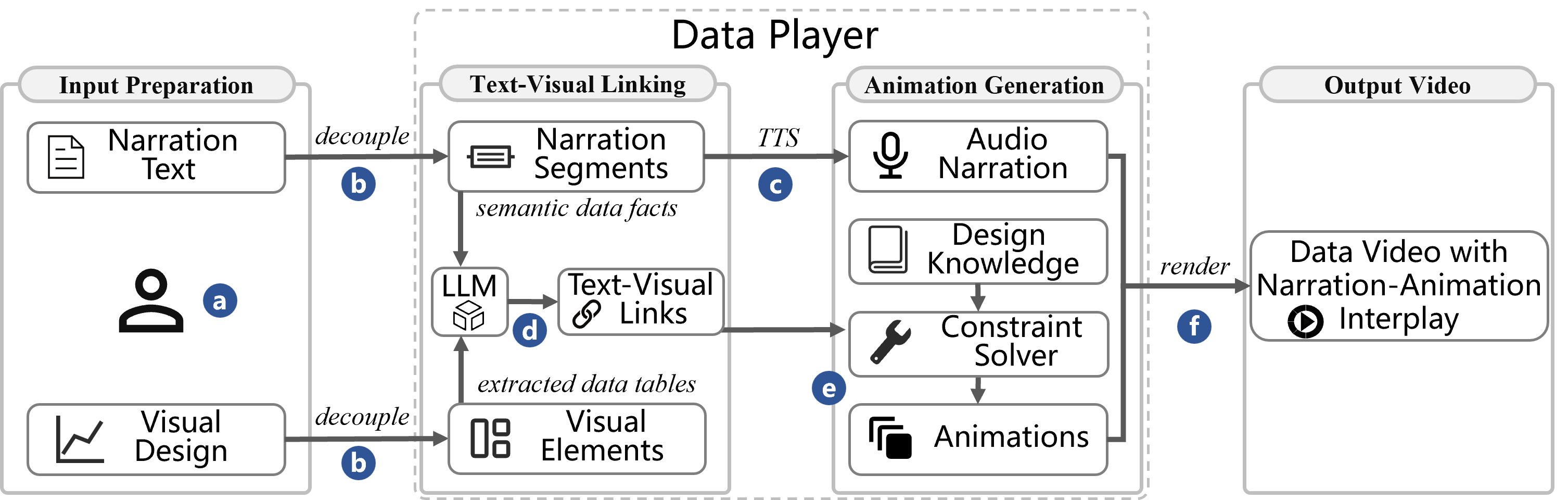}
\caption{The pipeline of automatic generation of data videos with narration-animation interplay.} 
\label{fig: pipeline}
\vspace{-4px}
\end{figure*}

\section{\toole}
In this section, we first introduce the conceptual model of data video with a set of design variables. Then we give an overview of \toole, and further discuss the two important modules: text-visual linking and animation sequence generation. 

\subsection{Data Video Modeling}\label{sec: Data Video Modeling}
Data videos consist of three design elements: visualizations, narrations, and animations. 
\begin{equation}
    video:= (visualization, narration, animation)
\end{equation}

\subsubsection{Visual Elements}\label{sec: visual elements}
Overall, visualizations are composed of a set of visual elements or element groups in a given static graphic.
\begin{equation}
    visualization := visual\;element | groups
\end{equation}

\begin{equation}
    group := visual\;elements
\end{equation}

The visual elements can be marks, axis, legends, annotations, etc. Each is defined as a tuple:
\begin{equation}
    visual\;element:= (id, type, data)
\end{equation}
\review{R1.4}\revise{where \emph{id} is the unique identifier, \emph{type} indicates the graphic shapes and visualization structure of the element, and \emph{data} is embedded in each visual element like dSVG in Canis~\cite{ge_canis_2020}. For each element, \emph{data} can be null and can also correspond to multiple data items.}

\subsubsection{Narration Entities}


Static narration text will be converted into audio speech, and each entity will be an audio unit with time; thus, a conceptual model of the narration text is:
\begin{equation}
    narration := narration\;entities
\end{equation}
\begin{equation}
   narration\;entity:= (audio, time)
\end{equation}
\begin{equation}
    time:= (start, duration)
\end{equation}
where \emph{start} refers to the start timestamp of an audio unit and \emph{duration} refers to the time span that an audio unit lasts.



\subsubsection{Animation Elements}\label{sec: Animation Elements}
The animation sequence applied in data videos is a series of animation \emph{units}. The animation is defined as: 
\begin{equation}
    animation := animation\;units
\end{equation}
\begin{equation}
   animation\;unit:= (visual\;elements, time, action, effect)
\end{equation}
where each animation unit targets a visual element or a group of elements and declares a start time and a duration. Additionally, the \emph{action} specifies which kind of animation \emph{effect} can be applied.

\subsection{Overview}
Prior research has pinpointed specific features of narration-animation interplay in high-quality data videos~\cite{wang2022investigating}.
Additionally, the formative study identified four common stages in the process of creating data videos, each of which requires considerable time and manual effort from users. 
We aim to automate the process of creating data videos with narration-animation interplay, making them more accessible and user-friendly for novice users. To this end, we design \tool that consists of two modules:
(1) \textit{Understand input narration text and visualization, and semantically link them.} Narration text frequently captures the central messages of data stories, incorporating data facts and insights associated with the visualization. The module automatically parses the text to extract the data facts presented in the narration. Further, it establishes connections between narration segments and visual elements within the visualization, which can be further leveraged to create audio and animations in the data video. (2) \textit{Recommend animation sequence and synchronize audio narration and visual animations.} Using visual cues corresponding to the words spoken in the narration is crucial in data video creation, and animations can make data more engaging and memorable for the viewer. To avoid any confusion or misleading the audience, 
the module automatically generates an appropriate animation sequence that serves the same purpose and intent as the storytelling, and synchronizes the audio narration with the visual animation, ensuring that the viewer receives the information clearly and coherently.

We propose an automatic pipeline to guide the design of \toole, as shown in \autoref{fig: pipeline}.
\review{R1.4}\revise{First, a static visualization and corresponding narration text are inputted in the form of Scalable Vector Graphics (SVGs) and plain text (a), respectively, so that they can be decoupled into multiple visual elements and narration segments (b).}
Then, Text-To-Speech (TTS) techniques are used to generate audio voiceovers and return timestamps of each word, which also act as the timeline of the data video (c).
Furthermore, the Large Language Model (LLM) is adopted to establish the semantic links between visual components (one or a group of graphic elements) and narrative entities (one or more words) based on the data facts to be told (d). 
\review{R2.2}\revise{To be specific, the linking module identifies the visual elements of the visualization inputs that can be animated, extracts data facts from them into tables, and links the table rows with semantic entities in the narration.}
After that, the animation generation module encodes collected design knowledge about narration-animation interplay into computational constraint programs and leverages the constraint solver to generate a suitable animation sequence with pre-designed animation presets based on the established text-visual links (e). 
Moreover, the module seeks to automatically organize animation sets in alignment with the generated audio timeline. It makes temporal decisions to allocate using constraint-based programming. 
As a result, a sequence of audio-animation packs is specified, which are further rendered into the data video (f).



\subsection{Text-Visual Linking}\label{sec: text-visual linking}
To generate data videos with narration-animation interplay, it is crucial to understand the narration text and its relations with the visual elements. We propose an LLM-based approach, shown in \autoref{fig: linking}, to generate these semantic links for animation. By extracting visual candidates that can be semantically linked in the visualization, LLM is then used to match these candidates with relevant narration segments.
We illustrate this below with an example of a 15-day PM2.5 value in Beijing.

\subsubsection{Data Extraction}

As described in Section~\ref{sec: visual elements}, each visual element has a \emph{data} property that contains semantic information. 
To effectively organize and utilize this information, we transfer the semantics into data tables and group elements based on the data items they contain.

\begin{figure*}[t]
	\setlength{\abovecaptionskip}{0.2cm}
	\setlength{\belowcaptionskip}{-0.1cm}
\centering
\includegraphics[width=0.97\linewidth]{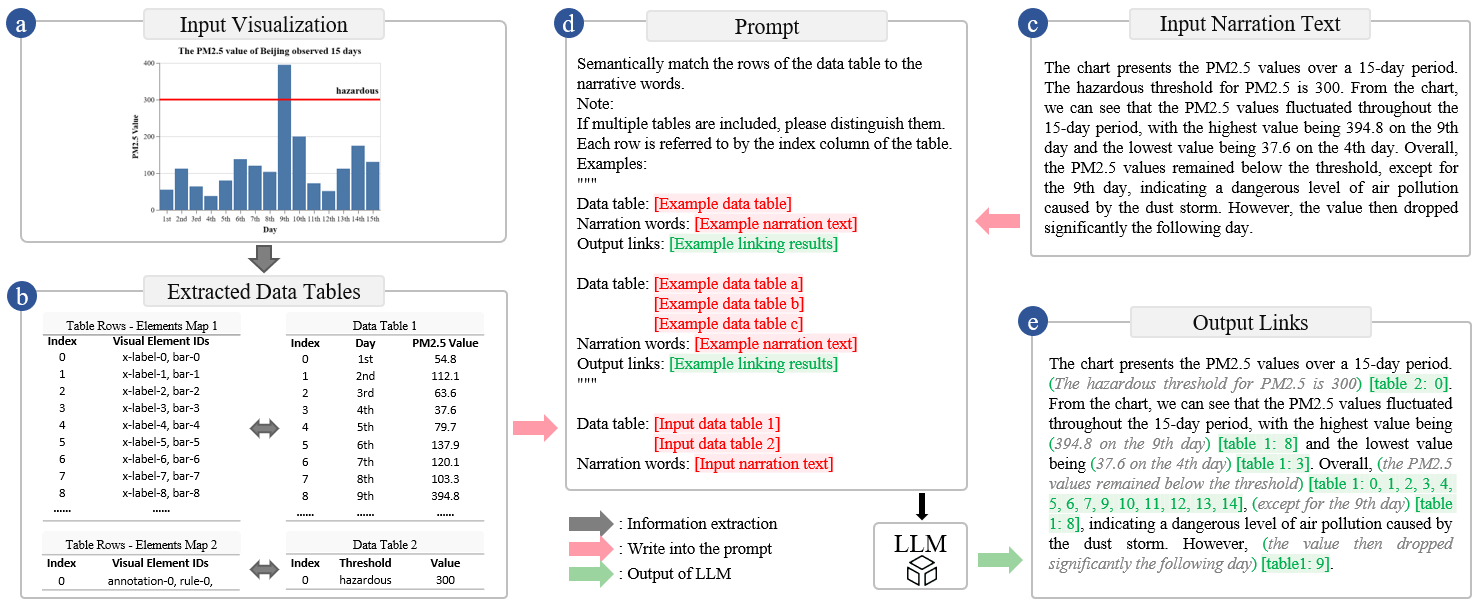}
\vspace{5px}
\caption{A walkthrough example illustrating the text-visual linking workflow. 
Data information behind the visualization (a) is extracted into data tables (b). 
The LLM with appropriate prompts (d) accepts the narration text (c) and data tables (b) as input and outputs semantic links between data table rows and narration segments (e).} 
\label{fig: linking}
\vspace{-4px}
\end{figure*}

Visual candidates in the visualization can be divided into two categories: basic graphical representations (e.g., marks, axes, legends, etc.) and annotations~\cite{Ren} (\autoref{fig: linking}-a).
First, as demonstrated in \autoref{fig: linking}-b, our method consolidates the data information in the SVG into a basic data table, which includes all values represented by graphical marks and axes. We also maintain a map that correlates each data table row to the corresponding visual elements. For example, the first row of data (Day: 1st, PM2.5 Value: 54.8) corresponds to both a bar mark (id is bar-0) and an x-label (id is x-label-0). 
While the data table captures most of the information present in the static chart, there may be missing information, particularly in regard to annotations, which play an important role in information communication. 
Referring to the annotation design space proposed by Ren et al.~\cite{Ren}, we divide annotations into text, graphics (including shapes and images), and their combinations. 
These elements contain valuable semantic information, such as the text (``hazardous'') and the red rule annotation in \autoref{fig: linking}-a, both of which express information about the hazardous threshold (300). Therefore, we extract a separate data table and group the corresponding elements.
Overall, each input data visualization will derive one data table for the chart marks and optionally one or more data tables for the annotations.

\begin{figure*}[t]
	\setlength{\abovecaptionskip}{0.2cm}
	\setlength{\belowcaptionskip}{-0.1cm}
\centering
\includegraphics[width=\linewidth]{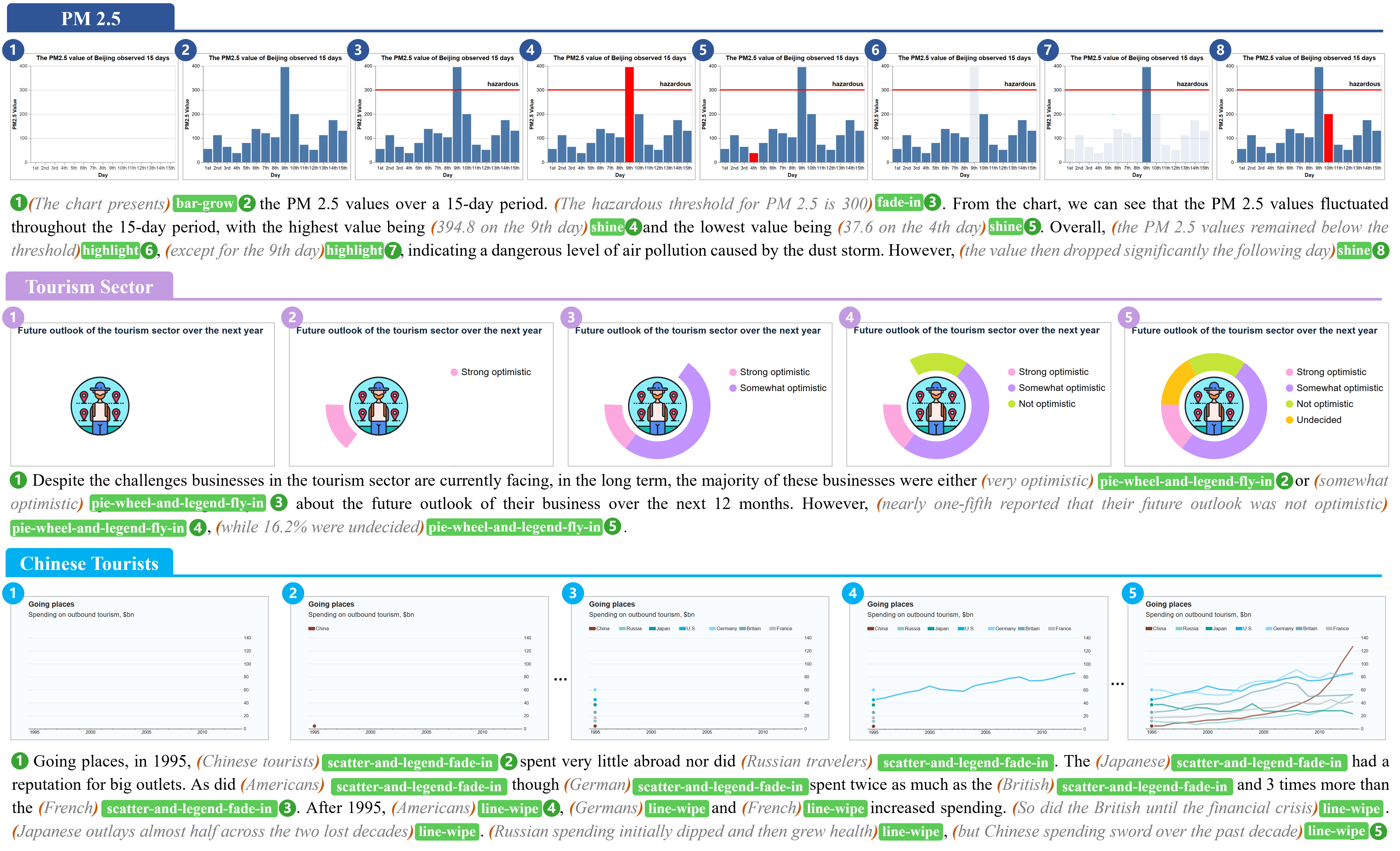}
\caption{Automatic-generated data videos by \toole. Each example includes a sequence of animation. The animation will be triggered when the audio narration reaches the corresponding segment. The snapshot images show the effect after the animation has been triggered.} 
\label{fig: example}
\vspace{-4px}
\end{figure*}

\subsubsection{Linking by an LLM}\label{sec: linking}
We have extracted semantics from static visualization graphics, as well as the mapping relationship between semantics and visual elements. To link narration segments (one or more words) and visual components (one or a group of graphical elements), the next step is to detect the occurrences of similar entities of these semantics in narration text. 

We further model this linking problem as a matching problem, semantically matching narration segments (\autoref{fig: linking}-c) with data table rows (\autoref{fig: linking}-b), and then mapping the data table rows to visual elements.
Specifically, we leverage the powerful natural language understanding ability of LLM (We use the OpenAI gpt-3.5-turbo model in our work.) to link the two perspectives, as shown in \autoref{fig: linking}-d. The prompt engineering aims to ask the LLM to accept data tables and narration words as input, and output semantic links as ``(\textit{narration segments})[table $x$: $R_i,...$]'', where $x$ is the table index and $R_i$ is the data table row index (see \autoref{fig: linking}-b).
\review{S.2}\revise{Inspired by existing successful prompt engineering experiences~\cite{Prompt, Liua},} the prompt includes few-shot pre-defined examples for a better illustration of our task.
The LLM output is shown in \autoref{fig: linking}-e. \review{R4.3}\revise{Finally, having links between table rows and narration segments in hand, we further obtain text-visual links with the help of mapping relationship between table rows and visual elements.}
In addition, we also fine-tune and deduplicate the text-visual links to avoid unnecessary animations in the subsequent steps.

\review{R3.4}\revise{
Our approach has several advantages over existing works about establishing references between text and charts~\cite{latif2021kori, lai_automatic_2020, Badam2019}.
First, we formulate the problem of text-visual linking into a problem of matching data table rows and narration words, which allows us to capture the semantic relationships between text and visuals more effectively. For instance, interpreting the phrase ``the following day'' requires an understanding of the temporal context of the data table, which is difficult to achieve by traditional rule-based linking methods. Second, we leverage LLMs, state-of-the-art language models, to perform similarity matching between data table rows and narration words with high accuracy due to larger knowledge support and better natural language understanding ability compared to prior NLP packages. 
However, traditional methods outperform the LLM-based method in terms of real-time performance.
}



\subsection{Animation Sequence Generation}
%

In this subsection, we introduce the animation recommendation module, which encodes collected high-level design knowledge into low-level constraints to automatically generate a suitable animation sequence.

\subsubsection{Animation Modeling}\label{sec: animation effects}
\review{R1.3}\revise{According to the formative study and existing literature~\cite{wang2022investigating},} designers concern with three types of semantics to implement appropriate animation effects. Specifically, they distinguish the semantic beginning and end of the narrative description of one data fact, as well as the emphasis intent in it and the information complement to the chart. Therefore, based on the definition in Section \ref{sec: Animation Elements}, we specify three animation actions:
``enter'' animations are applied when an object appears on the canvas and ``exit'' animations are applied when an object disappears from the canvas, while ``emphasis'' animations are applied to draw attention to an object that is already on the canvas. Each action includes several commonly seen changes in visual channels (e.g., ``fade'', ``grow'', ``zoom'', etc.). The timing and duration of the animations can also be adjusted to suit specific needs.

\subsubsection{Constraint Encoding}\label{sec: Constraint Encoding}
Prior work has demonstrated methods to generate designs from a set of design constraints~\cite{moritz_formalizing_2019, swearngin_scout_2020, chi_automatic_2020}, which motivates us to formulate narration-animation interplay design into a Constraint-Satisfaction Problem (CSP). 
In detail, We model the design elements discussed in Section~\ref{sec: Data Video Modeling} (i.e., visualization, narration, and animation) with encoded variables. Each variable has a domain. For example, animation actions include “enter”, “exit”, and “emphasize” (Section \ref{sec: animation effects}), and each action corresponds to a set of animation effects. The variable domains of visual elements and narration entities are derived from the generated text-visual links (Section \ref{sec: linking}).
To generate the animation sequence, \tool assigns concrete values to specific variables and leverages the CSP solver to explore numerous combination alternatives in the large search space.
Specifically, we encode high-level design knowledge summarized from the formative study and existing literature~\cite{wang2022investigating,amini_hooked_2018,shi2021communicating,tversky_animation_2002, heer_animated_2007} as computational low-level constraints. All constraints are formalized as equations and fed into the Z3 \cite{DeMoura2008} CSP solver. The solver outputs suitable animations. For instance, the animation sequence specified for the example in \autoref{fig: linking} is shown in \autoref{fig: example} (top). \review{S.4}\revise{Ultimately, the audio narration and visual animations are rendered into a dynamic data video (.mp4 format) with narration-animation interplay with the FFmpeg multimedia framework~\cite{FFmpeg}.}
The low-level constraint encoding is detailed as follows:



First, to ensure basic visual design quality, we use the established text-visual links to generate visual structure and data facts constraints, matching textual and visual entities and grouping visual elements. Specifically, we design \texttt{linking} constraints that allow only the visual elements that are linked to specific narration segments to be animated. 
The \texttt{integrity} constraints ensure that all elements involved in text-visual links need to be animated. We design \texttt{group} constraints to group the visual elements that are related to data facts in the text-visual links. Meanwhile, we design the \texttt{association} constraints to ensure that if one element is linked to narration, itself and other elements in the same data group can be animated. In addition, our \texttt{consistency} constraints specify that elements from different groups that are visually consistent should apply the same animation.


Second, we encode sets of temporal constraints to time-align animations and narration. Each group of constraints specifies how different elements of the data video should be timed and arranged on the timeline according to their type, effect, and relation to the narration. 
First, narration text inherently contains a chronological relationship between words. We further use Microsoft Azure Text-to-Speech services~\cite{TTS} to automatically generate audio narration and obtain the timestamps of each word in the audio, which also acts as the timeline to arrange animation effects applied to the visual elements. 
On this basis, we encode a \texttt{duration} constraint to determine that the animation effects are triggered by the onset of the first word in each linked narration segment, and last for the duration of the corresponding text span in the audio. The last frame of the previous animation will be retained for the time period when no animation is applied. 
The \texttt{conflict} constraints enforce the inherent logical order of animation actions. For example, visual elements can only be emphasized or disappeared after they appear, and elements cannot be emphasized after they disappear. 
And \texttt{on screen} constraints determine when an element appears or disappears from the canvas based on the ``$on\_screen$'' variable assigned to each element. If ``$on\_screen$'' is true at time $t$, then the corresponding element is visible at that time. Otherwise, it is hidden. By assigning different values of ``$on\_screen$'' to each element at different timestamps, we can create a table that shows which elements are on the canvas at any given moment. 
The table can help us control the animation actions to avoid overlapping or conflicting movements.
For instance, visual elements that have an enter animation applied will not appear on the canvas until the animation is triggered. Elements that have an exit animation applied will disappear from the canvas after the animation. Elements that do not have any animation applied will appear on the canvas by default.
In addition, a set of \texttt{order} constraints defines an optional logical sequence of elements such as background, title, axis and data items and the \texttt{synchronization} constraints ensure that elements in the same data group activate together. 



Third, different animations can produce different effects and serve different purposes. To align the semantic intents between the linked narration segments and animated visual elements, we encode a variety of constraints to assign appropriate animations from the pre-defined library to visual elements based on the data facts being presented, the visual structure, and the desired audience engagement.
We also specify constraints on animated annotations to avoid messing up the canvas.
\review{R1.5}\revise{Furthermore, we define a series of implicit mappings with priorities between the involved visualization structures and the appropriate animation combinations based on long-term practical experience.} These mappings are effective for defining the animations of the elements within a group. For instance, in a pie chart, the sector and its corresponding legend elements (e.g., symbol and label) are usually bound into a group. So we define a new animation called ``pie-wheel-and-legend-fly-in'', which means that the pie chart's sector will wheel clockwise and the legend-related elements will fly in at the same time, as shown in \autoref{fig: example} (middle). As a result, we can apply only one animation to multiple elements, avoiding specifying animations for each element individually.
On this basis, we define an objective function to minimize the number of animations used: $\min \sum_{i=1}^{m} A_i$, where $A_i$ is the number of animations applied to the $i$-th text-visual link and $m$ is the number of text-visual links. This function ensures that the module uses our predefined animation combinations as much as possible to maintain narrative coherence.

\subsubsection{Animation Presets}
A comprehensive library of animation effects for each action and mappings between semantics and effects can enable a wide range of designs. However, constructing such a large-scale library requires significant development costs. \review{R4.1}\revise{Thus, we utilize a small set of pre-designed animation effects based on the GSAP animation platform~\cite{GSAP} for different actions as a technology probe and proof-of-concept to explore our main research concern~\cite{wonderflow}.}
For instance, ``fade-in'' and ``wipe'' for entrance, ``zoom-out'' and ``fade-out'' for exiting, and ``shine'' and ``change-color'' for emphasis, etc. Additionally, depending on different chart types and element orientations, the configurations of one animation effect are adjusted, such as the ``grow'' effect for bar marks, ``wipe" for lines, and ``wheel'' for circular marks.
In the future, we will explore more vivid animations to enrich the library. 

%% file: sections/user-study.tex
\section{Evaluation}
To evaluate the liveliness of data videos generated by \toole, we (1) built an example gallery from real-world data storytelling practices, (2) conducted a user study to compare automatic-generated videos with those created by novices and designers, and (3) performed expert interviews to further understand the difference between automatic-generated data videos and human-composed ones.

\subsection{Example Gallery}
To demonstrate the expressiveness of the automatic approach, we generate a variety of example data videos based on a set of public design files. 
These examples cover a wide range of visualisation types (e.g., bar, pie, line, etc.) and narrative themes (e.g., PM2.5, tourism, stock price, tax payment, etc.). \autoref{fig: teaser} and \autoref{fig: example} show a subset of cases, more data video examples can be found in \url{https://datavideos.github.io/Data_Player/}.

\subsection{User Study}
In this study, we aim to understand the quality of data videos produced by \toole, by comparing them with data videos produced by novices and designers. 

\begin{figure*}[h]
	\setlength{\abovecaptionskip}{0.2cm}
	\setlength{\belowcaptionskip}{-0.1cm}
\centering
\includegraphics[width=\linewidth]{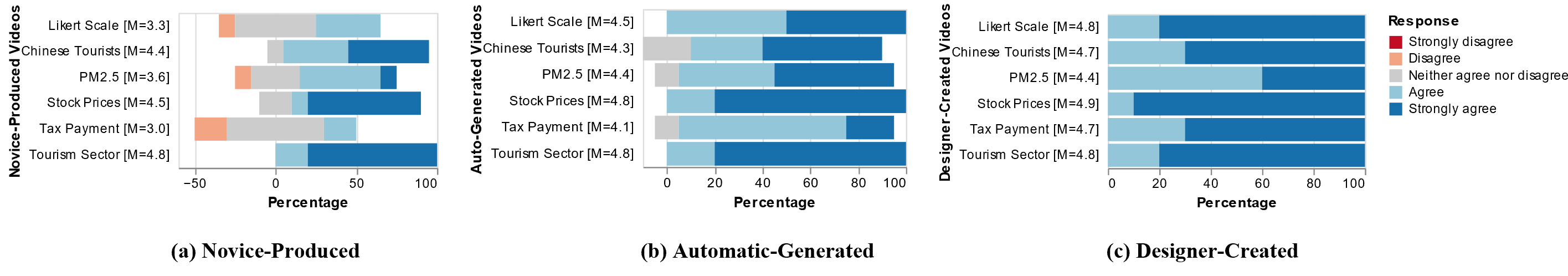}
\caption{User study results with a 5-point Likert scale. From left to right are novice-produced, automatic-generated, and designer-created videos.} 
\label{fig: user study}
\vspace{-8px}
\end{figure*}

\subsubsection{Dataset}\label{sec: dataset}
To prepare data videos for the user study, we collected six sets of static charts and their descriptions from real-world data storytelling practices, including 
a line chart with stroked point markers that shows spending on outbound tourism of Chinese tourists (short as ``Chinese Tourists'', \autoref{fig: example} (bottom)), 
a pie chart that describes the future outlook of the tourism sector over the next year (short as ``Tourism Sector'', \autoref{fig: example} (middle)), 
an annotated bar chart that depicts the PM 2.5 value of Beijing observed in 15 days (short as ``PM 2.5'', \autoref{fig: example} (top)), 
a stacked bar chart that describes America's tax system (short as ``Tax Payment'', \autoref{fig: teaser}),
a diverging stacked bar chart for sentiments towards a set of eight questions with a 5-point Likert scale (short as ``Likert Scale''), 
and a multi-series line chart that shows the stock prices for five high-tech companies (short as ``Stock Prices'').


\subsubsection{Procedure}\label{sec: procedure}
\textbf{Pre-experimental preparation.} 
We first invited four participants to manually create videos, including two novices and two professional designers. The novices were unfamiliar with video creation and only had experience using visualization to present data insights and using MS PowerPoint to create animations. The two designers' daily work involved using professional video creation tools like Adobe After Effects. The two designers also participated in our prior formative study (Section \ref{sec: formative study}). Based on the static visualizations and descriptions collected from the internet (Section \ref{sec: dataset}), each participant was asked to create three videos.
In order to control the conditions of comparison, we implemented an interactive authoring tool~\cite{wonderflow} that allows participants to manually specify the links between narration segments and visual elements, apply appropriate animations, generate audio from text, fine-tune the timeline, and preview the data video. They can ask us any questions they have during the manual creation process to ensure they can master the tools without compromising the quality of their creative data videos. Finally, we confirmed with them that the data videos they produced were representative of their level of design.
For each piece of material, we also obtained an automatically generated version using \toole. In total, we collected six novice-composed data videos, six designer-created data videos, and six automatic-generated videos. 

\textbf{During experiment.} 
We recruited another 10 participants (denoted as P1 to P10) who have data analysis and visualization needs in their daily work, including data analysts, ML researchers, software engineers, and visualization researchers. Each participant viewed six sets of videos, each set including three versions: novice-created, automatic-generated, and designer-composed. 
The sequence of the three videos and the six sets was shuffled. We also provided the textual descriptions that were used to create the data videos.
The participants can watch the video repeatedly as they like and were asked to rate the overall quality of each video in terms of expressiveness and liveliness using a 5-point Likert scale. They were encouraged to leave reasons for the decisions and speak of any comments on any aspect of the videos. 
After all data videos have been viewed and evaluated, we asked participants to identify the automatic-generated one in each set. The experiment lasted about 60 minutes.


\subsubsection{Results}

\textbf{Subjective Satisfaction:}
The results of the participants’ ratings are shown in \autoref{fig: user study}. From left to right are novice-produced, automatic-generated, and designer-created data videos. Overall, participants had a positive sentiment towards all the data videos. The majority of participants rated the videos as “agree” or “strongly agree”, with mean values all greater than or equal to 3.0.
In detail, the results showed that the videos automatically produced by \tool were generally well-received by the participants, with a mean score of 4.4 (\textit{SD}=0.70) for ``PM2.5'', 4.3 (\textit{SD}=0.82) for ``Chinese Tourists'', 4.1 (\textit{SD}=0.57) for ``Tax Payment'', 4.8 (\textit{SD}=0.42) for ``Stock Prices'', 4.5 (\textit{SD}=0.53) for ``Likert Scale'', and 4.8 (\textit{SD}=0.42) for ``Tourism Sector''. These scores were higher than the mean ratings given to the novice-produced data videos and comparable to the mean ratings given to the designer-created videos.
In terms of the specific topics of the data videos, the automated data videos received the highest mean ratings for ``PM2.5'' and ``Tourism Sector'', with means of 4.4 and 4.8, respectively. The designer-composed videos had slightly higher mean ratings than automatic-generated ones for other topics. However, the paired t-test results showed that there is no significant difference in ratings between automatic-generated and designer-created data videos for all topics. 

We also found that for visualizations with relatively simple structures and patterns, there is little variation in the user ratings of different types of videos. For example, all three types of videos received the same ratings for ``Tourism Sector'' (both 4.8) and the paired t-test results show that the difference is not significant between ratings for ``Chinese Tourists'' and ``Stock Prices''. However, when the chart structure is more complex (e.g., ``Tax Payment'' and ``Likert Scale''), especially when the narrative expresses some in-depth content, novices often encounter difficulties to tell the story more reasonably, and the designers' experience can help them deal with these situations better. The automatic algorithm was able to effectively communicate the information in a clear and engaging way, at a level between novices and designers.

Participants can identify the automatic-generated videos from the three versions with an average accuracy of only 31.7\%. 
This means that the automatic-generated videos were close to the human-composed ones. Moreover, we found that 45\% of the misjudged videos were actually composed by novices, and 23.3\% were actually made by designers. 
This suggests that the level of skill and creativity of the human composers also affected the perception of the evaluators.

\textbf{Feedback:}
All participants agreed that narration-animation interplay can enhance the efficiency and vividness of data insights communication compared to static forms. They also praised the data videos' expressiveness, liveliness, and overall quality, and noted that they would consider using the automatic method in their own work.
P2 said, ``\textit{the videos dynamically present information while ensuring completeness. I was so surprised that the wonderful data videos were generated automatically.}'' They also expressed a preference for some aspects that were focused on in our design knowledge. For example, all participants appreciated the use of animation in the right context. P4 also commented, ``\textit{the videos were visually consistent overall, as the same animations were used for similar visual structures.}''

We also learned some lessons about users' preferences. Some participants criticized the visual effects and style of the videos and hoped that the algorithm could better meet individualized needs. For example, P1 did not like the ``Bounce'' animation effect of bar marks, and P3 did not expect the ``Change Color'' effect to always be red. For legend, P3 preferred to have the graphical elements and their corresponding legends presented together in sequence so that the audience can get an immediate understanding of the visual elements. While P1 preferred to see the legend first to get a general impression of the context, and then the marks and the narration appear in sequence. 
This indicated that a user interface is needed to extend the existing automation pipeline and incorporate humans into the workflow.

\subsection{Expert Interview}
To further compare the difference between automatic-generated videos and human-composed ones, \review{S.6.2}\revise{we invited six experts to provide feedback through interviews, including four designers (denoted as D1 - D4), of whom D1 and D2 helped us create designer-composed videos in the user study (Section \ref{sec: procedure}),} and two visualization researchers (denoted as V1 and V2), who have more than five years of experience conducting visualization research and publishing visualization papers in major conferences (e.g., IEEE VIS and ACM CHI). 
Moreover, all of them joined our prior formative study (Section \ref{sec: formative study}) and provided valuable feedback.
They were asked to watch the six sets of data videos used in the user study, and they were informed of the respective versions in advance.
Then they need to provide specific feedback on comparing different versions. 
\review{S.6.2}\revise{In addition, they were also asked to comment on the method's strengths and weaknesses, its possible application scenarios, and the future outlook.}

Overall, all the participants agreed that all data videos were able to effectively interact between narration and animation.
When comparing the three data video versions, the participants generally thought that there was no very obvious difference between them overall. 
\review{S.6.3}\revise{
However, D1 and D2 found that humans (especially designers) did a better job fine-tuning the timeline. 
In the creation of data videos, the animation duration is dependent on the length of the narration segment in the text-visual link. This may result in animations that are too short, such as those corresponding to only one word, which can lead to user confusion, particularly for animations intended to emphasize certain points. They (D1 and D2) also noted that they spent a certain time previewing and iteratively refining the animations for the designer-created videos, including adjusting the trigger timestamp and duration of animations, so they can mitigate this issue. 
Participants (D1, D3, D4, V1, and V2) also pointed out the animation-intensive nature of \toole. D1 stated, ``\textit{the automatic algorithm feels like it wants to add animations to every sentence,}" while V2 agreed, ``\textit{this level of animation density can be tiring for viewers.}'' and V1 complemented,  ``\textit{I often found it hard to keep track of multiple animated visual elements at the same time, especially when they move or change in different ways.}''

The participants also provided valuable insights into the potential application scenarios and future improvement directions. 
All of them agreed that \tool is an effective tool for empowering novices to create data videos.
V1 and V2 also suggested that \tool can be used as a module integrated into other large systems to prototype more complex videos, automate slide design, etc. 
D3 commented that our automatic technique can be further extended to enable other creative ways of storytelling, such as scrollytelling ~\cite{seyser_scrollytelling_2018} and interactive data videos~\cite{hook_facts_2018}, which requires more interactive and exploratory experience.
D1 and D4 expected that professional video production software (e.g., Adobe After Effects) can integrate the animation recommendation module. 
Even though all participants praised the convenience of \toole, some of them (V1, V2, D2, D4) also suggested further automatically generating static visualizations and narration text for users. More importantly, during the process, the system should allow users to fine-tune the generated results at each stage of data video generation to provide an interactive human-in-the-loop experience.
}

%% file: sections/discussion.tex
\section{Discussion}
\textbf{Automation vs. Personalization.}
Creating narration-enriched data videos is a highly specialized and time-consuming task. \tool can help users automate this process based on the input static visualization and description. It enables rapid exploration of design alternatives, thus increasing efficiency for data insight presentation.
However, in our user study, we found that participants' personal preferences (e.g., animation effects and visual styles) may affect their ratings of data videos. These highly personalized needs are difficult to be thoroughly satisfied in a full-automatic algorithm~\cite{Wu2021d}. 
\review{R3.2}\revise{Subsequently, we plan to design a user interface and develop Human-AI collaboration methods~\cite{Li2023a}. 
First, the automated method can prototype data videos (including automatic generation of visualizations and corresponding narration text), and then users can further modify them with fine-grained control on the interface. Next, the system can allow users to input their preferences at different stages of data video generation, and progressively generate data videos.
Moreover, the system can provide various animation examples for users to select and adapt to their own designs~\cite{Shen2022b}, and the system can also automatically learn personalized needs from the user's interaction history through multi-modal interactive task learning~\cite{Ruoff2023}. 
Another interesting approach would be to help users maintain a personalized knowledge base. Based on existing professional design knowledge, users can continuously expand and update their personalized rules. }


\textbf{Application of LLMs.}
\review{R3.4}\revise{We applied LLMs to match narration segments and visual elements. Before this, we tried traditional NLP packages and BERT-based n-gram similarity matching schemes, which were somewhat mechanical and rigid.} Recently, LLMs (e.g., chatGPT and GPT 4) have demonstrated remarkable capabilities in generating and understanding natural language. After using LLMs, the matching module achieved better accuracy and flexibility. 
However, existing LLMs still have some inherent limitations, such as being not good at complex computational tasks, producing inconsistent outputs in different rounds, taking a long time for generation, and hallucination problems, etc. These factors affect the accuracy, timeliness, and practicality of our methods to some extent. But we believe that future research will address these issues.
In the future, we will mainly explore three directions based on LLMs: The first is how to better improve the accuracy of text-visual linking with LLMs, such as exploring more accurate prompts and integrating with other interactive tools~\cite{Wu2022d}. \review{R3.2}\revise{The second is to further expand the existing automatic pipeline based on LLMs, such as helping users interactively generate and modify narrations~\cite{Shen2023} and automatically create and update visualizations~\cite{vistalk} based on the data table. 
The third is to fit the use of LLMs within human-in-the-loop scenarios. For example, LLMs can extract insights from data. Users can also choose the insights of interest based on their analytic tasks~\cite{Shen2021}, and further leverage LLMs to automatically generate targeted 
narration text, visualizations, and chart annotations.
Furthermore, given some specific material (e.g., a visualization or a narration segment), LLMs can be used for similarity searching to obtain more material to aid storytelling~\cite{Li2022, Linenet}.
}


\textbf{The completeness of design knowledge.}
Design knowledge plays an important role in the narration-animation interplay design. In this paper, we have explored several key design constraints from the formative study and existing literature. However, it is important to note that these constraints are not exhaustive and we only consider them as a minimal set to assist data video creators and researchers in crafting narration-animation interplay for data videos. There may be other factors that have an impact on the narration-animation interplay in different contexts and domains (e.g., spatial alignment, visual complexity, and cognitive load). Therefore, we encourage further research to investigate more design guidelines to assist data video creators in creating more effective and engaging narration-animation interplay.
Additionally, the target function in our current animation recommendation module can be further improved. In the future, we can also consider setting hard (must be satisfied) and soft (will be penalized if not satisfied) constraints~\cite{moritz_formalizing_2019} to enable more flexible recommendations.


\textbf{Understanding emotions in the narration.}
Our automatic pipeline primarily focuses on the semantic matching of narration segments and visual elements, as well as the rationality of animations. We also use text-to-speech technology to automatically generate audio narration. This allows us to generate reasonable and vivid data videos, but we do not further understand the emotions from the narration.
In the future, it would be interesting to investigate the emotional design space in narration-animation interplay and express emotions in the narration with appropriate emotional animations, tones, and visual cues~\cite{lan2021kineticharts,Xie2023}. Additionally, in order to draw attention to significant events and changes in animation, it is important to use sound effects, which are also not considered in our pipeline. Furthermore, having background elements appear at the beginning can also serve to set the tone and mood of the video. For example, if the video is presenting visualization related to a serious topic, such as a disease outbreak, having a somber and serious title card at the beginning can help to establish the tone of the video.

\textbf{Limitations and Future Work.}
In our user study, we found that the quality of the input visualization and narration itself can influence users' judgments. In the future, we can use users' own materials to create videos and then ask them to compare the videos with their previous data presentation forms. 
\review{S.6.1}\revise{Another limitation of our study was the relatively small number of participants. It would be interesting to conduct a larger crowdsourced user study to further verify the quality of the automated data videos and to investigate the factors that contribute to their perceived quality.}
\review{R3.2}\revise{Additionally, \tool currently only supports data video generation with a single chart and structured data. In the future, it can be extended to more visualization scenarios (e.g., multiple charts or dashboard~\cite{Lin2022}, glyph~\cite{Ying2022}, and infographics~\cite{wang_animated_2021}) and data types (e.g., geographic~\cite{Li2023b}, graph~\cite{Shen2022}, and word cloud~\cite{Xie2023}) to tell more complex stories.}

%% file: sections/conclusion.tex
\section{Conclusion}
To streamline the complex process of crafting narration-animation interplay for data videos, this paper proposes \toole, 
which enables the automatic transformation of static text and visualizations into engaging data videos, enabling more novices to share their insights and research findings using data videos.
\tool leverages advanced LLMs to extract data facts and establish text-visual links, and uses constraint programming to recommend animations for the links and time-align audio narrations and visual animations. 
The results of the user study indicated that the data videos automatically produced by \tool were well-received by participants and comparable in quality to human-composed ones. 
We hope that the approach can help people with little video production experience quickly create high-quality data videos, and inspire future research about narration-animation interplay in data storytelling.